\documentclass{emulateapj}

\usepackage{graphics,graphicx}
\usepackage{natbib}
\def\apj{\rm ApJ}
\def\apjl{\rm ApJL}
\def\apjs{\rm ApJS}

\def\mnras{\rm MNRAS}
\def\nat{\rm Nature}

\newcommand{\msun}{M$_{\odot}$}

\shorttitle{MBH Seeds}
\shortauthors{Bellovary et al.}

\begin{document}

\title{The First Massive Black Hole Seeds and Their Hosts}

\author{Jillian Bellovary\altaffilmark{1}, Marta Volonteri
\altaffilmark{1}, Fabio Governato\altaffilmark{2}, Sijing
Shen\altaffilmark{3}, Thomas Quinn\altaffilmark{2}, James
Wadsley\altaffilmark{4}}

\altaffiltext{1}{Department of Astronomy, University of Michigan, Ann
Arbor, MI}

\altaffiltext{2}{Department of Astronomy, University of Washington,
Seattle, WA}

\altaffiltext{3}{Department of Astronomy, University of California Santa Cruz,
Santa Cruz, CA}

\altaffiltext{4}{Department of Physics and Astronomy, McMaster
University, Hamilton, ON, Canada}

\begin{abstract}

\noindent
We investigate the formation of the first massive black holes in high
redshift galaxies, with the goal of providing insights to which
galaxies do or do not host massive black holes.  We adopt a novel
approach to forming seed black holes in galaxy halos in cosmological
SPH+$N$-Body simulations.  The formation of massive black hole seeds
is dictated directly by the local gas density, temperature, and
metallicity, and motivated by physical models of massive black hole
formation.  We explore seed black hole populations as a function of
halo mass and redshift, and examine how varying the efficiency of
massive black hole seed formation affects the relationship between
black holes and their hosts.  Seed black holes tend to form in halos
with mass between $10^7$ and $10^9$ \msun, and the formation rate is
suppressed around $z = 5$ due to the diffusion of metals throughout
the intergalactic medium.  We find that the time of massive black hole
formation and the occupation fraction of black holes are a function of
the host halo mass.  By $z = 5$, halos with mass $M_{\rm halo} >
3\times10^{9}$ \msun~ host massive black holes regardless of the
efficiency of seed formation, while the occupation fraction for
smaller halos increases with black hole formation efficiency.  Our
simulations explain why massive black holes are found in some
bulgeless and dwarf galaxies, but we also predict that their
occurrence becomes rarer and rarer in low-mass systems.

\end{abstract}


\section{Introduction}

Massive black holes (MBHs) are commonly found in massive galaxies with
a significant bulge component \citep{Gehren84}, but recent discoveries
show that they can also be found in bulgeless disk galaxies
\citep{Filippenko03,Shields08} and dwarfs \citep{Barth04,Reines11}.
It is then unclear whether bulges are related to MBH formation
processes, what fraction of low-mass galaxies may host MBHs, and
whether MBHs in these galaxies have an important role in galaxy
evolution.  In this paper we present a novel approach to seed MBH
formation based on the properties of metal-free gas which explains the
existence of MBHs in dwarf galaxies \citep[e.g.,][]{Reines11}. Our
simulations, however, suggest that MBHs become infrequent in low--mass
galaxies, and that a minimum galaxy mass exists below which MBHs
become progressively uncommon.

MBHs must have originated from moderately massive ``seed'' black holes
in order to grow to a billion solar masses by $z\simeq 6$
\citep{Haiman01}.  These seeds must form at high redshifts ($z \sim
15-30$) and grow rapidly in order to reproduce the observed
distribution of high redshift quasars \citep{Volonteri03}.
Seeds may be the remnants of Population III stars, which form with
extremely low metallicity and thus have unique properties.  It is
commonly speculated that Population III stars have a top-heavy IMF,
with masses ranging from 100 - 1000 \msun~
\citep[e.g.][]{Couchman86,Abel02,Bromm04}.  Any zero-metallicity star
with a mass greater than $\sim 260$ \msun~ will leave a $\sim 100$
\msun~ black hole behind \citep{Bond84,Heger02}.

Another theory involves the direct collapse of very metal-poor,
low-angular momentum gas via dynamical instabilities
\citep{Oh02,Loeb94,Eisenstein95,Koushiappas04,Begelman06,Lodato06}.
If enough gas is funneled into the center of a local overdensity, it
may collapse to form a black hole with mass $10^4 - 10^6$ \msun~
\citep{Begelman06,Lodato06,Begelman08}.  This process may happen later
than Population III star formation, because halos must be larger to
host such massive inflow. Efficient gas collapse is more likely to
occur in massive halos with virial temperatures $T_{\rm vir} > 10^4$K
under metal-free conditions where the formation of H$_2$ is inhibited
by a UV background \citep{BrommLoeb2003}, and cooling is dominated by
atomic hydrogen. In such halos, fragmentation is suppressed, cooling
proceeds gradually, and the gaseous component can cool and participate
in MBH formation before it is turned into stars.  These halos may need
to exist in regions of ultracritical UV radiation in order to form
MBHs by direct collapse \citep{Dijkstra08,Shang10}, since the average
estimated UV background may not be sufficient to prevent some
Population III stars from forming in halos of this size
\citep{Johnson08}.

While the processes which may lead to MBH seed formation have been
modeled cosmologically \citep{Wise08,Regan09}, it is not yet possible
to do so in simulations involving volumes larger than a few Mpc$^3$.
Previous large-scale simulations have incorporated seed black hole
formation in a simplistic way.  \citet{Sijacki07}, \citet{DiMatteo08}
and \citet{Booth09} all employ similar methods, which involves running
an on-the-fly halo finder on the simulation as it evolves, and
planting seed MBHs in particular halos by hand.  The halos are chosen
based on a mass threshold (generally $M_{\rm halo} \sim 10^{10}$
\msun), and seeds are planted if there is not already a black hole
present.  There is no metallicity criterion, so seed MBHs can form at
any redshift if a halo meets the eligibility criteria.  The seeds are
placed at the center of the halo and generally fixed there throughout
the remainder of the simulation \citep{Booth09}.

However, there is no physical motivation for a halo mass threshold for
MBH placement of $10^{10}$ \msun, and in fact seeds likely form in
halos of much lower mass (see \S \ref{sect:history}).  This method
also prevents more than one seed from forming per halo, which may be
unrealistic - halos may experience the formation of multiple
Population III stars if fragmentation occurs
\citep{Turk09,Stacy10,Greif11,Clark11}.  Fixing the black hole at the
halo center is similarly inadvisable - dynamical effects such as
galaxy mergers or gravitational recoil may cause a black hole to
temporarily vacate the exact center of its galaxy, but such a
circumstance is prohibited in these models.

The formalism of seed MBH formation we describe in this paper is a
more sophisticated and realistic model because it makes no assumptions
about the underlying halo properties.  Our method relies only on the
prospective MBH's local environment to determine when and where the
MBH forms, including a requirement for zero-metallicity gas.  Our
scenario is broadly consistent with either proposed MBH formation
mechanism, and also allows MBHs to evolve dynamically in a realistic
way. It is crucial to study how variations in halo mass affect the
frequency of formation when examining simulated seed MBH populations,
and so we include in our sample a variety of halo masses in order to
present a more coherent picture of high-$z$ seed formation and
evolution as a function of cosmic environment.  With this technique,
we investigate where and when MBH seeds form at high redshift, and
discuss the implications for galaxies at $z = 0$.

\section{The Simulations}

We employ the SPH+$N$-Body Tree code Gasoline
\citep{Stadel01,Wadsley04}, which has been shown to produce realistic
galaxies in cosmological simulations \citep[e.g.][]{Governato10,
Stinson10,Pontzen10,Oh10,Brooks11}.  Gasoline includes a physically
motivated prescription for star formation as well as a recipe for the
formation and evolution of MBH seeds (described in \S
\ref{sect:form}).  We use a Kroupa IMF \citep{Kroupa} and a WMAP3
cosmology \citep{WMAP3}.  We use a supernova feedback model which
incorporates the Sedov solution to the blastwave equations
\citep[see][]{Stinson06}, and set the blastwave energy to $E_{SN} =
10^{51}$ ergs.  We implement a uniform ionizing radiation background
with an onset at redshift $z = 9$ \citep{Haardt96}.  Our simulations
include cooling through metal lines as well as a model for turbulent
metal diffusion \citep{Shen10}; however, we do not including cooling
via molecular hydrogen in the simulations presented here.

To probe a large parameter space, we employ a suite of simulations
with different halo masses to test seed MBH formation.  It is crucial
to study how variations in halo mass affect the frequency of formation
when examining simulated seed MBH populations.  We have chosen a set
of three fiducial halos selected from a uniform resolution, 50 Mpc
volume and resimulated at high resolution using the volume
renormalization technique \citep{Katz93}.  We sample the region of
interest at high resolution, and then sample more coarsely as the
distance from the chosen object increases.  This technique results in
a large dynamic range, where we can capture the detailed physics of
galaxy formation on small scales in the region of interest, and also
the large-scale tidal torques from structures several Mpc away.  The
three chosen halos correspond broadly to a {\it low-mass} disk galaxy
($h603$), a {\it Milky Way-mass} disk galaxy ($h258$), and a {\it
massive elliptical} galaxy ($hz1$) at $z = 0$. We present the details
of our simulations in Table \ref{table:sims}.  Our simulations are
extremely high resolution, with gas particles having masses between $2
- 9 \times 10^4$ \msun, and a force resultion of 173 - 260 pc,
depending on the simulation.  The mass ratio of gas and dark matter
particles (column 4) is of order unity, which reduces the effects due
to two-body scattering and is critical for keeping MBHs in the centers
of their galaxies.

\begin{deluxetable*}{lccccccc}
  \tablecolumns{8} 
  \tablewidth{0pc}
  \setlength{\tabcolsep}{0.015in} 
  \tablecaption{Simulation Properties
    \label{table:sims}
  }
  \tablehead{
    \colhead{Simulation}& \colhead{Halo Mass}&\colhead{Gas particle }&\colhead{gas/DM} & \colhead{Softening }& \colhead{$\chi_{\rm seed}$ } & \colhead{\# BHs} & \colhead{$n_{min}$ }\\
    & at $z=5$ (\msun) & mass (\msun) & ratio  & (kpc)  & & @ $z = 5$  & (amu/cm$^{-3}$) }

  \startdata
  h603  & $8.48 \times 10^9$ & 26676 & 1.69 & 0.173 & 0.05, 0.1, 0.3, 0.5 & 33, 44, 51, 49 & 10\\
  h258 & $2.92 \times 10^{10}$ & 26676 & 0.713 & 0.173 & 0.05, 0.1, 0.3, 0.5 & 39, 56, 76, 84& 10\\
  hz1  & $5.88 \times 10^{11}$ & 90031 & 0.713 & 0.260 & 0.05, 0.1, 0.3, 0.5 & 166, 240, 465, 504 & 2.5\\
  \enddata
\end{deluxetable*}

Since we are interested in the epoch of seed MBH formation, we have
run our simulations to $z = 5$ rather than $z = 0$, in order to
maximize resolution while using only modest compuational resources.
We use the Amiga Halo Finder \citep{Gill04,Knollmann09} to identify
galaxy halos based on the overdensity criterion for a flat universe
\citep{Gross97}.  For each simulated region, we analyze MBH
populations for the primary halo as well as every satellite with at
least 64 particles.   In the case of $h603$ there are 5370 total
halos in our analysis; for $h258$ there are 5170; for $hz1$ there are
2160.

\section{Seed Formation and Evolution}\label{sect:form}

Since the actual physical process of MBH seed formation is
unresolvable in cosmological simulations, we have developed a model
which is broadly consistent with both of the proposed seed formation
scenarios.  The common thread between the proposed scenarios is an
ability for gas to collapse into a large central mass, which requires
a zero or near-zero metallicity (though see \citet{Mayer10} for an
alternate mechanism).  Because our star formation prescription is
already based on similar physics (i.e. cold, dense, collapsing gas
results in the formation of star particles), we use this prescription
with the additional criterion of zero metallicity to form seed MBHs.

\subsection{MBH Formation}

Our current star formation recipe is described in detail in
\citet{Stinson06}, and we summarize it here.  For a star to form in
our simulations, its parent gas particle must meet several criteria.
Primarily, the gas density must be greater than the threshold density
for star formation, $n_{min}$.  In simulations with force resolution
of $\lesssim$ 100 pc, high density peaks can be resolved, mimicking
star formation regions in giant molecular clouds \citep{Governato10}.
In this instance, one can use a realistic value of $n_{min}$ = 100 amu
cm$^{-3}$.  However, at lower resolution we cannot properly resolve
these high density peaks, and in order to match observed relations
such as the Kennicutt-Schmidt law \citep{Kennicutt89} and the
Tully-Fisher relation, we use a lower value (related to the mass and
spatial resolution) for $n_{min}$ (2.5 - 10 amu cm$^{-3}$).  These
values are presented in Table \ref{table:sims} for each simulation in
our study.  In addition to the density criterion, the gas temperature
$T$ must be less than a fiducial temperature $T_{max}$, which we set
to $1.5 \times 10^4$ K.  Star formation efficiency is governed by the
free parameter, $c^*$, which we set to 0.1 \citep{Governato10} to
match the observed relations mentioned above.  If a gas particle meets
all of the relevant criteria, the probability it will form a star is
given by

\begin{equation}
p = \frac{m_{\rm gas}}{m_{\rm star}} (1 - e^{c^*\Delta t/t_{\rm form}})
\end{equation}

where $m_{\rm star}$ and $m_{\rm gas}$ are the star and gas particle
masses, $t_{\rm form}$ is the dynamical time for the gas particle, and
$\Delta t$ is the time between star formation episodes, which we set
to 1 Myr.

For a seed MBH to form, all of the criteria for star formation must be
met in addition to the criterion of zero metallicity. If a gas
particle meets the criteria to form a star and probabilistically is
able to do so, it then has an additional probability to instead form a
MBH.  The probability to form a MBH seed is given by the above
expression multiplied by the newly introduced parameter $\chi_{\rm
seed}$, which sets the approximate number of seed MBHs which will form
in a given simulation.  In this paper we investigate how varying this
parameter affects MBH populations in high redshift galaxies with a
range of masses, and explore values of $\chi_{\rm seed}$ between 0.05
and 0.5.  This parameter range is motivated by models predicting the
efficiency of seed MBH progenitors.  For example, according to
\citet{Lodato06} the distribution of halos which might host
direct-collapse MBHs varies between 4 - 35\% for a reasonable range of
halo spins, virial temperatures, and the Toomre parameter
$Q$. Additionally, the Population III star IMF reported by
\citet{Tan10} has a mean mass of 250 \msun, which is roughly the mass
where one expects a MBH remnant to form; therefore $\sim 50\%$ of
Population III stars may form MBH seeds.  Thus we believe our
$\chi_{\rm seed}$ range of 0.05 to 0.5 samples the parameter space of
current theoretical predictions.

When seed MBHs do form, they acquire the mass of their parent gas
particle.  Due to resolution limitations, we cannot form MBH seeds
with Population III remnant masses ($\sim 100$ \msun), but one can
imagine a scenario in which Pop III seeds form at high $z$ and grow
through mergers and accretion to the higher masses we employ in our
simulations \citep{Li2007}.  Our seed mass is therefore broadly
consistent with either seed formation scenario.  MBHs are allowed to
merge if they become close together in space (within two softening
lengths) and have low relative velocities.  Specifically, they must
fulfill the criterion $\frac{1}{2} \Delta \vec{v}^2 < \Delta \vec{a}
\cdot \Delta \vec{r}$, where $\Delta \vec{v}$ and $\Delta \vec{a}$ are
the differences in velocity and acceleration of the two black holes,
and $\Delta \vec{r}$ is the distance between them.  In this study, we
do not enable MBHs to grow through gas accretion, nor do we include
any type of feedback from MBHs.  We have chosen to omit these aspects
of MBH physics to gain a robust lower limit on MBH seed populations.
We verified that accretion and feedback do not affect our results by
performing a simulation of $h258$ with the inclusion of these
processes (assuming Bondi-Hoyle accretion and a feedback efficiency of
1\%, see \citet{Bellovary10} for more details), and found that there
were no significant differences in MBH growth or star formation
history with their inclusion.  At the high redshifts we are probing
here, such MBH activity is only efficient in the largest halos, which
are rare in the majority of the simulations we present in this work.

\subsection{MBH Formation History}\label{sect:history}

\begin{figure*}[htb]
\begin{center}
\includegraphics[scale=0.6]{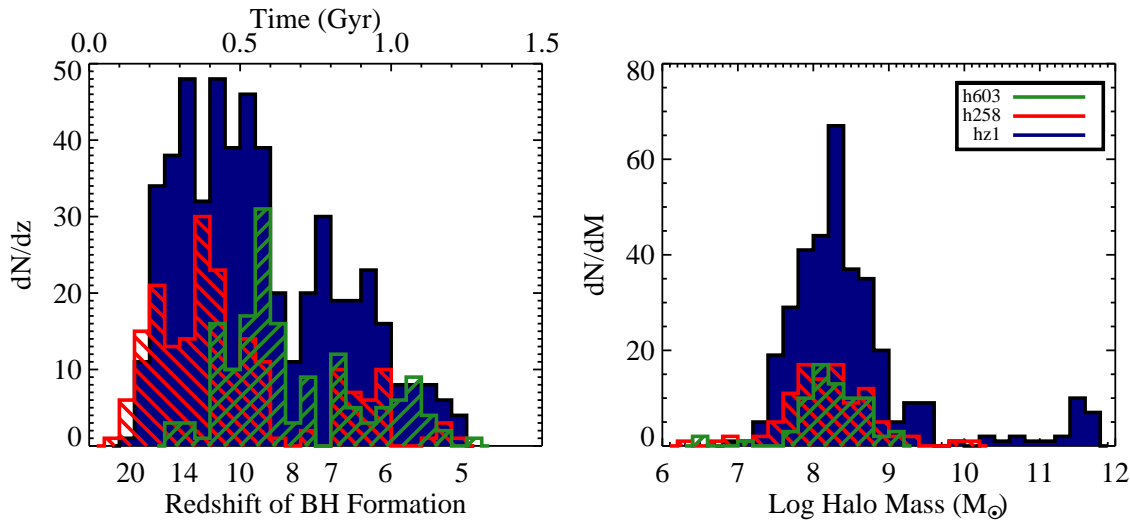}
\caption{ {\em Left:} Formation history for $\chi_{\rm seed} =
  0.1$, for MBHs in all resolved halos in simulations $h603$
  (green), $h258$ (red), and $hz1$ (purple). {\em Right:}
  Mass of the halos at the time the MBH formed for $\chi_{\rm seed}
  = 0.1$.
  \label{fig:compare}
}

\end{center}
\end{figure*}

A detailed understanding of the formation history of MBHs, as well as
the host halo mass at the time of MBH formation, is essential for
interpreting our results.  In Figure~\ref{fig:compare} (left panel) we
show the formation history of seed MBHs for every resolved halo in
each galaxy for $\chi_{\rm seed} = 0.1$.  Changing $\chi_{\rm seed}$
does not qualitatively change the MBH formation history, it simply
results in a different number of MBH seeds formed (higher values of
$\chi_{\rm seed}$ result in increased seed formation and consequently
an increased MBH-MBH merger rate - see Table \ref{table:sims} for the
number of MBHs in each simulation at $z = 5$).  Comparing simulations
$h603$ (green histogram) and $h258$ (red histogram), we see the clear
trend of an earlier onset of MBH formation with larger overdensities.
The peaks in MBH formation are offset by 200 - 400 Myr for simulations
$h603$ and $h258$ due to this effect of cosmic bias \citep{Peebles80}.
This effect is not evident between simulations $h258$ and $hz1$
(purple histogram) due to the stochastic nature of cosmic large-scale
structure.  However, though simulations $hz1$ and $h258$ begin forming
stars and MBHs at the same time, these objects form at a much more
rapid rate in $hz1$ due to the overdense nature of the selected
region.  The dip at $z = 9$ for each simulation is a result of the
onset of the UV background at that time, when a subsantial amount of
cool gas is heated above $T_{max}$.  In all galaxies the rate of seed
formation dwindles around $z = 5$, because the majority of halos have
experienced supernova explosions and have been metal-enriched, which
in turn quenches seed MBH formation.  At this point, MBH seeds can
only form in areas which have not yet experienced local star
formation, which are generally restricted to the outskirts of the
simulation.

\begin{figure}[htb]
\begin{center}
\includegraphics[scale=0.5]{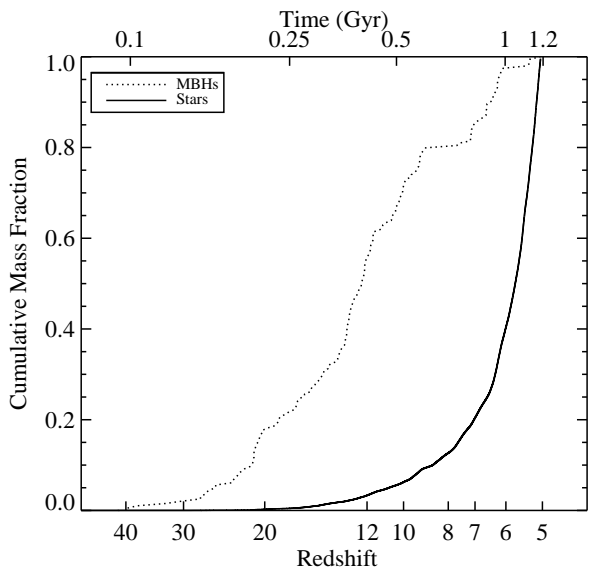}

\caption{ Cumulative mass fraction for stars (solid line) and MBHs
(dotted line) for simulation $h258$ with $\chi_{seed}$ = 0.1.
  \label{fig:form}
}

\end{center}
\end{figure}

The connection between star formation, metal pollution and MBH
formation is exemplified in Figure \ref{fig:form}.  Here we show the
cumulative mass fraction of stars and MBHs for all of the galaxies in
simulation $h258$ ($\chi_{seed} = 0.1$).  The bulk of the MBHs form
while the star formation rate is still quite low.  MBH formation
tapers off at $z = 5$, at a time when the star formation rate is
increasing rapidly.  These effects are due to the metallicity
criterion for seed formation; MBHs form efficiently only until metals
have diffused through the interstellar gas, at which point MBH
formation is truncated fairly quickly.

We next examine the mass of a halo at the time a black hole seed forms
within it.  Figure~\ref{fig:compare} (right panel), shows that in our
simulations seed MBHs form in halos which are less massive than the
$M_{\rm halo} = 10^{10}$ \msun~ assumed in previous works. Our
predicted halo mass range is between $10^7$ and $10^9$ \msun, with a
peak around $10^8$ \msun, and does not vary with $\chi_{\rm seed}$.
(Again we show only $\chi_{\rm seed} = 0.1$, the other values of
$\chi_{\rm seed}$ give nearly identical results.) This halo mass is
consistent with that in which Population III stars are expected to
form \citep[$> 10^6$ \msun,][]{Couchman86,Barkana01}, and also with
the halo mass expected for seed MBH formation via direct collapse
scenario \citep[$\sim 10^7$
\msun,][]{Koushiappas04,Begelman06,Lodato06}.  This agreement is
simply a consequence of the physics of gas cooling in virialized
halos; though we do not include prescriptions for molecular hydrogen
cooling or Population III star formation, we find good agreement with
predictions of the first collapsing metal-free gas clouds.  The halo
mass at time of MBH formation is slightly dependent on the resolution
of the simulation in question, in that the distribution shifts toward
smaller masses with higher resolution, but this effect is minimal and
we do not expect our results to differ strongly if we move to yet
higher resolution.  Also note that because more than one MBH is
allowed to form in a halo, a particular halo may be represented more
than once in the right panel of Figure 1.

The halo mass at the time of MBH formation is somewhat dependent on
the cosmic overdensity, in that MBH formation (and star formation in
general) begins earlier in volumes with larger density perturbations.
This phenomenon leads to the earliest onsets of star formation
occurring in halos which are slightly smaller than those in less
overdense regions.  For the galaxy $hz1$, the distribution of halos is
actually bimodal, with a strong peak around $M_{\rm halo} \sim 10^8$
\msun, and a secondary peak at large halo mass.  The MBHs which form
in large halos do so at late times (after the primary burst of MBH
formation) in the outskirts of the primary halo (between 20-50
physical kpc from the center) in pockets of very low metallicity.
These MBH formation events may be a resolution effect, as they
may be occuring in pressure-confined clouds as reported in
\citet{Kaufmann06}; with higher resolution (and a corresponding higher
density threshold for star/MBH formation) these objects may cease to
exist.  We plan to explore this effect further with high resolution
tests.  However, since these objects form in low density regions,
they will exist as ``wandering'' MBHs in the halo and never inspiral
to the center to merge with the primary.

Allowing MBH seeds to form in halos with $M_{\rm halo} < 10^{10}$
\msun~ is vital for capturing their full evolution.  While several of
these seeds will eventually become central MBHs in large halos at $z =
0$, many may have other fates.  For example, when small halos hosting
MBH seeds merge with larger ones, they may undergo tidal stripping,
resulting in a ``wandering'' black hole in the larger galaxy halo
\citep{Bellovary10}.  If the satellite halo is not tidally destroyed,
it still may have been stripped of enough baryons to quench the growth
of its MBH, resulting in a dwarf galaxy hosting an intermediate mass
black hole which may be near its initial seed mass.

\section{MBH - Halo Occupation Fraction}

The frequency at which MBHs occupy halos is a fundamental measure of
MBH seed formation efficiency.  There are constraints on this quantity
in the local universe, albeit weak ones; for example, nuclear activity
due to MBHs has been detected in 32\% of the late-type galaxies in the
Virgo Cluster, exclusively in galaxies with mass $M_{\rm halo} >
10^{10}$ \msun~ \citep{Decarli07}.  For early-type galaxies in Virgo,
nuclear activity exists in 3-44\% of galaxies with mass less than
$10^{10}$ \msun, and 49 - 87\% of galaxies with mass greater than
$10^{10}$ \msun~ \citep{Gallo08}.  Such estimates put a lower limit on
the MBH occupation fraction of galaxies in this cluster, since
inactive black holes are unlikely to be observed.  Observations at
higher redshifts are more challenging, though data at $z \sim 1$ from
DEEP2 and AEGIS may help provide some constraints
\citep[see][]{Yan11}, and the synergy of {\em JWST} and {\em ALMA}
will more robustly probe occupation fractions in distant galaxies.

 Theoretical constraints of the MBH occupation fraction are also
quite weak.  Semi-analytic models show that the occupation fraction
varies depending on which method of MBH formation is used
\citep{VanWassenhove10} or how efficiently seeds are formed
\citep{Volonteri08,Tanaka09}.  Only a few percent of high redshift halos need
host MBHs in order to reproduce the $z = 0$ occupation fraction
\citep{Menou01}.  However, high-redshift occupation fractions anywhere
from 10\% to 100\% can reproduce the observed quasar luminosity
function and SMBH mass function, depending on parameters such as
radiative efficiency and quasar duty cycle \citep{Lippai09}.  Clearly
much uncertainty remains regarding the high redshift MBH occupation
fraction, and we take this opportunity to explore how halo mass and seed
formation efficiency affect this quantity.

\begin{figure}[htb]
\begin{center}

\includegraphics[scale=0.53]{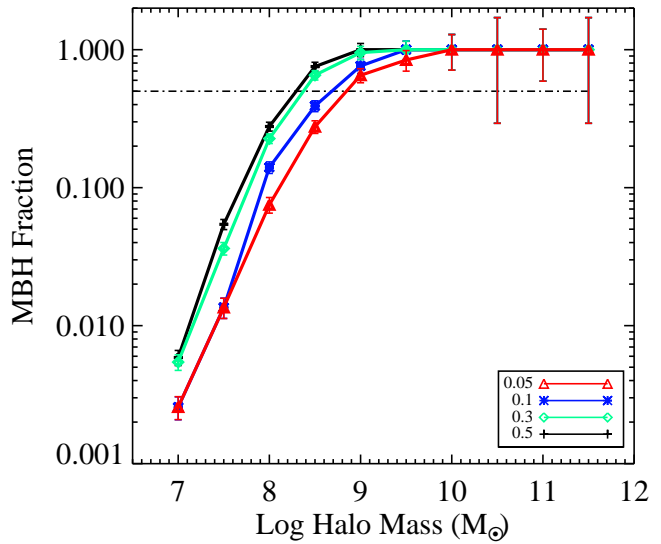}

\caption{The MBH-halo occupation fraction for a given halo mass, for
the parent galaxy and subhalos of all three simulations at $z = 5$.
Colored lines and symbols represent simulations with different values
of $\chi_{\rm seed}$ (red triangles: $\chi_{\rm seed} = 0.05$, blue
asterisks: $\chi_{\rm seed} = 0.1$, green diamonds: $\chi_{\rm seed} =
0.3$,black crosses: $\chi_{\rm seed} = 0.5$).  Error bars represent
uncertainty due to Poisson statistics.  The horizontal dot-dashed line
marks where the probability that a galaxy hosts a MBH is 50\%,
corresponding to halo masses between $10^8$ and $10^9$\msun.
\label{fig:occupation}
 }

\end{center}
\end{figure}

In Figure \ref{fig:occupation} we show the MBH-halo occupation
fraction vs. halo mass for the four values of $\chi_{\rm seed}$. Here
we include the parent galaxy and subhalos of all three simulations
together at $z = 5$.  Halos with virial masses above log$(M_{\rm
halo}) \sim 9.5$ always host a MBH seed, regardless of the value of
seed formation efficiency.  Even in the lowest efficiency case, MBH
seeds form in the regions of earliest star formation, which tend to be
the halos which become the most massive later on in every simulation.
The most massive halos have also experienced the greatest number of
mergers, which further populates them with MBH seeds brought in by
satellites.  Thus, our model predicts that halos with masses greater
than $M_{\rm halo} \sim 10^{9}$ \msun~ will be extremely likely to
host MBH seeds, even if the formation efficiency of such seeds is
small.  This result is consistent with observations of the local
universe, where MBHs are found to occupy halos above a similar mass
threshold with high likelihood \citep{Ferrarese06,Wehner06}.

To estimate the fate of MBHs at later cosmic times, we have run a
lower-resolution simulation of galaxy $h258$ to $z = 0$ and traced the
mass evolution of halos. Halos with mass between $10^7$ \msun~ and
$10^{10}$ \msun~ at $z = 5$ can remain in the same mass range at $z=0$,
be stripped, or become incorporated in the main halo.  In this particular
case, $\sim$ 30\% of halos with $M_{\rm halo} > 10^{9}$ \msun~ -- that
host MBHs already at $z = 5$ -- are stripped during their evolution,
ending up as halos with $z = 0$ masses of $10^8 - 10^{10}$ \msun.  While
one zoomed-in simulation cannot give us broad statistics on the
evolution of all $z = 5$ halos, we can deduce that while a large
number of high redshift MBH hosts undergo hierarchical merging and
settle in massive galaxies, a non-negligible fraction have more
quiescent merger histories and do not grow substantially, and make up
today's population of low-mass galaxies which may harbor MBHs.

On the other hand, the occupation fraction of halos with masses less
than $M_{\rm halo} \sim 10^{9}$ \msun~ is sensitive to the physics of
MBH formation, via $\chi_{\rm seed}$.  For example, in galaxies $h258$
and $hz1$, halos with mass $M \sim 10^9$ \msun~ are nearly 100\%
likely to host a MBH seed if $\chi_{\rm seed} = 0.5$, but only 10\%
likely when $\chi_{\rm seed} = 0.05$.  The dependence of occupation
fraction with $\chi_{\rm seed}$ becomes weaker in galaxy $h603$,
however.  We suggest that in the larger halos, metals diffuse
throughout the simulation more efficiently (due to increased star
formation and possible ejection of metals from galaxies) which
suppresses MBH formation in their satellites.  In the isolated
low-mass disk galaxy, however, the relative lack of metal diffusion
allows MBHs to continue forming in these halos at a time when MBH
formation would have been truncated in a larger overdensity.  In this
study we have aimed for high resolution with a sample of field
galaxies; for a simulated uniform volume, including voids and high
density regions (which more completely samples halos in the universe)
our results may be slightly different.  However, a test study using a
small (6 Mpc) uniform volume gives results consistent with our
findings here.

\section{Summary}

We have undertaken a study of the formation and evolution of seed MBHs
using cosmological simulations, including a unique and physically
motivated recipe for seed MBH formation.  In our simulations, MBH
formation depends solely on the local properties of the surrounding
gas (i.e. density, temperature, metallicity) rather than on global
halo properties.  We can thus model the evolution of MBH seeds within
their halos in a cosmological framework in a fully self-consistent
way.  

We find that MBHs form in a burst during the onset of the earliest
star formation, but the formation rate tapers off by $z = 5$ due to
the diffusion of metals throughout the intergalactic medium.  MBH
formation is concentrated in halos with masses between $10^7$ and
$10^9$ \msun, consistent with predictions of seed supermassive black
hole formation
\citep{Couchman86,Barkana01,Koushiappas04,Begelman06,Lodato06}.  MBHs
which exist in small halos at high redshift may contribute to the $z =
0$ population of ``wandering'' black holes in massive galaxies, or
they may appear as intermediate mass black holes in low mass and/or
bulgeless galaxies \citep{Greene07} if their hosts have quiescent
merger histories.

We find that the time of black hole formation and the occupation
fraction of black holes are a function of the host halo mass. Large
halos form MBHs earlier and they are more likely to host a MBH.  An
observational determination the MBH-halo occupation fraction for halos
of mass $M_{\rm halo} < 10^{9}$ \msun~ would be a strong constraint on
the true formation efficiency of MBH seeds \citep[see
also][]{Volonteri08b,Volonteri08,VanWassenhove10}.  In a forthcoming
paper we will investigate the co-evolution of MBHs and galaxies at
later cosmic times, as well as derive both the occupation fraction and
the active fraction of MBHs in galaxies. The latter is a direct
observational constraint; a preliminary study of the active fraction
at $z = 1$ is already possible now with data from DEEP2 and AEGIS,
while local high sensitivity studies are being carried for both field
galaxies (Miller et al. in prep) and in the Virgo Cluster
\citep{Gallo2010}.

It may be possible that some MBHs could form at even lower redshifts
in halos near cosmic voids.  Since these areas have below-average
densities, they will have experienced far less star formation and thus
will have a lower metallicity than their high-density counterparts.
Such events have been predicted for Population III stars at redshifts
from $2 < z < 6$ \citep{Jimenez06,Tornatore07,Trenti09}.  Searches for
seed MBH formation in high-redshift void galaxies may provide unique
observational clues regarding the origins of supermassive black holes.

\acknowledgements

Simulations were run using computer resources and technical support
from NAS. MV acknowledges support from SAO Award TM1-12007X and NASA
awards ATP NNX10AC84G and NNX07AH22G.  FG acknowledges support from a
NSF grant AST-0607819 and NASA ATP NNX08AG84G.  JB and TQ acknowledge
support from NASA Grant NNX07AH03G.  The authors also thank Kayhan
G\"{u}ltekin, Brendan Miller, Michele Trenti, and the anonymous
referee for providing insights which improved the paper.


\end{document}